\documentstyle[epsfig,longtable]{aipproc}
\newcommand{\beq}{\begin{equation}}
\newcommand{\eeq}{\end{equation}}
\newcommand{\beqa}{\begin{eqnarray}}
\newcommand{\eeqa}{\end{eqnarray}}

\newcommand{\nuf}{$\nu$FMSR}
\input{epsf}

\title{The use of neutrino beams from muon storage rings
\footnote{Invited plenary talks given at
23rd Johns Hopkins Workshop on Current Problems in 
Particle Theory: Neutrinos in the Next Millennium, 
Baltimore, MD, 10-12 June 1999, and
The International Conference on Physics Potential and 
Development of $\mu^+ \mu^-$ Colliders,  
San Francisco, CA, 15-17 December 1999.}  
}

\author{Alexey A. Petrov}
\address{Department of Physics and Astronomy\\
Johns Hopkins University\\
Baltimore, MD 21218}

\begin{document}
\date{}
\maketitle

\begin{picture}(0,0)
\put(320,200){JHU--TIPAC--200002}
\end{picture}

\begin{abstract}
I give a brief overview of the physics potential of
short baseline experiments at
neutrino factories, i.e. facilities providing
high energy and high intensity neutrino beams, like the one 
planned to be built in connection with the proposed 
high energy muon storage ring. These facilities
would offer a unique opportunity to perform new precision studies 
of QCD and electroweak interactions. New types of measurements, such 
as studies of gluon density of the nucleon via charmonium production and 
extractions of $V_{cb}$ and $V_{ub}$ CKM matrix elements, will become 
possible. Interesting new physics scenarios can also be explored.
\end{abstract}

\section{Motivation}

The purpose of this talk is to overview 
physics goals for the short baseline experiments utilizing 
high energy and high intensity neutrino beams.
These include standard model electroweak 
physics, novel tests of QCD, and rare processes
sensitive to physics beyond the standard model (SM).

The standard model electroweak parameters that are conventionally 
measured in neutrino experiments are $\sin^2 \theta_W$ 
and the Cabibbo-Kobayashi-Maskawa (CKM) quark mixing matrix 
elements $V_{cd}$ and $V_{cs}$. Given the intense high energy 
beam of neutrinos these measurements will certainly yield new
precise values for these quantities. In addition, completely new
measurements, like precision studies of $V_{ub}$ and $V_{cb}$
CKM matrix elements, will be possible. On the QCD side, neutrino-nucleon 
interactions are potentially the best probes of various
valence parton distribution functions, both unpolarized and 
polarized, as well as the strong coupling constant
$\alpha_s$. One can also study various
non-perturbative parameters, such as fragmentation functions.
Finally, there are interesting new physics scenarios that can be 
tested in neutrino interactions. These include supersymmetric extensions 
of the standard model with broken ${\cal R}$ parity (or any models 
with leptoquarks), new heavy neutral leptons or gauge bosons, etc.

A proposed muon storage ring should provide a highly collimated, 
high-intensity $\nu$ beam from muons decaying in the 
accelerator tunnel. The neutrino spectra can be easily  
calculated~\cite{Bigietal}; for instance, the $\nu_\mu$ energy 
spectrum is given by $d N_{\nu_\mu} /dx \simeq 6 x^2 - 4 x^3$ with
$x=2 E_\nu'/m_\mu$ being the normalized neutrino energy in the $\mu$
rest frame. $E_\nu'$ is easily related to the neutrino energy in the lab
frame, $E_\nu = x E_\mu (1+ \cos \theta)/2$, where $\theta$ is a neutrino
angle in the muon rest frame. Another advantage of this facility
is that for sufficiently high muon energies the neutrinos are
produced in thin pencil-like beams with an opening half-angle
$\theta_\nu \simeq m_\mu/E_\mu$.

\section{QCD and Electroweak studies}
\subsection{Measurements of $\sin^2\theta_W$} 

One of the most important parameters of the standard model 
is the weak mixing angle $\theta_W$ which represents the angle of 
rotation from the ``gauge'' basis to the ``physical'' basis 
where the mass matrix of the gauge $Z$ boson and the photon 
is diagonal. One of the many possible definitions is
the on-shell definition of $\sin^2\theta_W$
\begin{equation}
\sin^2\theta_W^{os} \equiv  1 - \frac{M_W^2}{M_Z^2},
\end{equation}
In neutrino-nucleon interactions $\sin^2\theta_W$ can be extracted 
using the Llewellyn Smith~\cite{LS83} or Pascos-Wolfenstein 
relations~\cite{PW73}. These methods involve measuring three 
total cross sections and forming three ratios,
\begin{equation}
R_\nu = \frac{\sigma(\nu N \to \nu X)}
{\sigma(\nu N \to \mu^- X)},~
R_{\bar \nu} = \frac{\sigma(\bar \nu N \to \bar \nu X)}
{\sigma(\bar \nu N \to \mu^+ X)},~
r = \frac{\sigma(\nu N \to \mu^+ X)}
{\sigma(\nu N \to \mu^- X)}.
\end{equation}
In the approach of Llewellyn Smith, these can be combined 
to obtain $\sin^2 \theta_W$:
\begin{eqnarray}
R_\nu &=& \frac{1}{2} - \sin^2 \theta_W +
\frac{5}{9}(1+r)\sin^4 \theta_W + C_\nu 
\nonumber \\
R_{\bar \nu} &=& \frac{1}{2} - \sin^2 \theta_W +
\frac{5}{9}(1+r^{-1})\sin^4 \theta_W + C_{\bar \nu},
\end{eqnarray}
where $C_\nu$ and $C_{\bar \nu}$ represent known QCD and electroweak 
corrections. Alternatively, a Pascos-Wolfenstein construction can be
used to extract $\sin^2 \theta_W$:
\begin{eqnarray} \label{PW}
R^\pm &=& \frac{R_\nu \pm r R_{\bar \nu}}{1 \pm r},
~
R^- = \frac{1}{2} - \sin^2 \theta_W +
\frac{C_\nu - r C_{\bar \nu}}{1 \pm r}
\nonumber \\
R^+ &=& \frac{1}{2} - \sin^2 \theta_W +
\frac{10}{9}\sin^2 \theta_W + 
\frac{C_\nu + r C_{\bar \nu}}{1 \pm r}.
\end{eqnarray}
This method is actually ``cleaner'' as the 
QCD and electroweak corrections partially cancel out in 
Eq.~(\ref{PW}). These relations are now used to extract 
$\sin^2 \theta_W$ by CCFR/NuTeV collaboration and will be 
used again at \nuf.

In addition to the methods described above, intense neutrino 
beams from the muon storage ring (\nuf) should allow for another
measurement of $\sin^2 \theta_W$, which involves neutrino-electron 
scattering. This method is theoretically ``cleaner'', as
it involves scattering of two leptons. This measurement 
involve investigation of four neutrino-electron elastic
cross sections $\nu_i (\bar \nu_i) e^- \to \nu_i (\bar \nu_i) e^-$
for $i=e, \mu$. The involved cross section are much smaller
then the corresponding DIS cross sections described above, but
theoretical clearness of this process and much improved 
neutrino beam intensity makes this measurement a realistic 
possibility. Of course, future determinations of $\sin^2 \theta_W$
from \nuf~ should be comparable or better than the projected 
result of the SLAC E158 Moller scattering experiment, i.e.
should measure $\sin \theta_W$ with relative accuracy 
of better then $a~few~\times 10^{-4}$. Preliminary 
studies~\cite{Bigietal} show that it is quite realistic.
\subsection{CKM and quark densities}

Extraction of the matrix elements of the 
Cabibbo-Kobayashi-Maskawa quark mixing matrix is
one of the outstanding challenges in phenomenology
of the standard model. It is most likely that Nature has 
chosen only three generations of quarks, so
\begin{eqnarray}
V_{CKM} = S_L^{u \dagger} S_L^d =
\left( \matrix{ V_{ud} & V_{us} & V_{ub}\cr
V_{cd} & V_{cs} & V_{cb}\cr
V_{td} & V_{ts} & V_{tb}\cr} \right).
\end{eqnarray}
Currently, the {\it charm} quark sector of CKM is best studied 
in $\nu N$ charged current interactions, where neutrino interacts 
with valence and sea quarks of the nucleon. For example, the CCFR 
collaboration has provided a direct measurement of 
$|V_{cd}| = 0.232^{+0.017}_{-0.019}$. Independent 
knowledge of the strange sea quark density should also provide 
an independent measurement of $|V_{cs}|$ as well. In the framework 
of a ``naive'' parton model,
\begin{eqnarray} \label{GePo}
\frac{d^2 \sigma (\nu N \to \mu^+ \mu^-X)}{d\xi dy} &=&
\frac{G_F^2 M E_\nu}{\pi}
\{
\left[
\xi u(\xi) + \xi d(\xi) \right] 
\left|V_{cd}\right|^2 
\nonumber \\
&+& 2 \xi s(\xi) \left|V_{cs}\right|^2
\}\left[1-\frac{m_c^2}{2 M E_\nu \xi} \right] D(z) B_c,
\end{eqnarray}
where $\xi = x(1+m_c^2/Q^2)$ in a ``slow rescaling'' model of
Georgi and Politzer~\cite{GePo76} and $D(z)$ is a charm fragmentation
function. Thus, measuring $d^2 \sigma$ at different values of $\xi$
provides an independent measurement of $|V_{cd}|$ and $|V_{cs}|$, if
quark densities are known well enough~\cite{Conrad:1997ne}. Otherwise, 
a multiparameter fit can be performed to determine both $q(x)$ and CKM 
matrix elements.

Even though in the above discussion a parton model was used, 
a problem of semiinclusive single particle production can be addressed 
model-independently using the formalism of perturbative QCD factorization
theorems. In this framework, the essential problem of having a heavy quark 
in the final state is the fact that its mass brings an additional 
scale to the problem at hand.
The presence of this scale might affect theoretical predictions by inducing
large logarithms involving $m_Q$, which have to be resummed in order for
perturbative expansion to make sense. A practical recipe for such resummation 
is provided by the prescription of Aivazis {\it et. al.} (ACOT)~\cite{ACOT}. 

A future measurement utilizing high intensity neutrino beams would
provide accurate determinations of various $q(x)$. It is interesting to 
note that future neutrino factories would be able to study production 
of the $b$-flavored mesons, which would allow for an accurate determination 
of the intrinsic charm content of the nucleon. If an $x$-dependent high 
statistics measurement of $b$-quark production becomes available, an 
independent determination of $|V_{ub}|$ and $|V_{cb}|$ CKM matrix elements 
will be possible as well~\cite{Bigietal}.

Of course, heavy quark production is not the only way of studying the
nucleon structure. Quark densities are usually measured in DIS-type 
experiments. These measurements are naturally performed in neutrino-nucleon 
interactions. Here, \nuf~ offers a tantalizing possibility to measure 
parity-violating
{\it polarized} nucleon structure functions. These measurements were considered 
hopeless in $\nu N$ experiments due to the enormous technical difficulties in 
polarizing heavy targets, the only possible targets for neutrino 
accelerator experiments if sufficient statistics is expected.
At \nuf~, light targets (like $H_2$ or $D_2$), which are relatively easy
polarized, can be used due to the large density of neutrino beam.
A number of unique measurements (such as the measurement of 
parity-violating polarized structure function of neutron)
is possible~\cite{Bigietal}.

\subsection{Charmonium production and gluon density}

High intensity neutrino beams would also allow studies of
{\it charmonium} production, a sensitive probe of the gluon
distribution function in the nucleon. Contrary to the open-flavor 
meson production, the production of charmonium states can be 
described in a model-independent fashion using the factorization
theorems of Non-Relativistic QCD (NRQCD), 
\begin{equation}
\sigma(A + B \to H_{c \bar c} +X) = \sum_n \frac{F_n}{m_c^{d_n-4}}\,
\langle 0 | {\cal O}^H_n | 0 \rangle,
\end{equation}
which separates short-distance physics, represented by the 
coefficients $F_n$ (which might be sensitive to various parton
distribution functions) from the long-distance physics,
represented by the NRQCD matrix elements
\begin{equation}
\langle 0 | {\cal O}^H_n | 0 \rangle \,=\,
\sum_X \sum_{m_J}\,
\langle 0 | {\cal K}_n | H_{m_J} + X \rangle\,
\langle H_{m_J} + X | {\cal K}^\prime_n | 0 \rangle,
\end{equation}
and determine the probabilities of charm quarks produced in the 
various angular momentum and  color (singlet and octet) states 
by action of NRQCD operators ${\cal K}^{(\prime)}_n$
to evolve into a physical charmonium state, like a $J/\psi$.
At the moment, these matrix elements cannot be computed
model-independently. However, they are {\it universal} 
(i.e. process-independent), 
so they can be extracted from other experiments.
Clearly, $J/\psi$ produced in sufficiently high numbers can be used to
study gluon distribution function in the wide range of 
$x$~\cite{Petrov:1999fm}.

A major advantage of using the neutrino beam is that, at leading order in
$\alpha_s$, the spin structure of the $\nu Z$ coupling selects a certain 
combination of octet operators.
The largest contribution is from the one with the quantum
numbers~$\,^3S_1^{(8)}$. The differential cross section was 
calculated in~\cite{Petrov:1999fm}:
\begin{eqnarray}
\frac{d\sigma\left(s,Q^2\right)}{dQ^2}&=&
\frac{\pi^2\alpha^2\alpha_s}{3\sin^42\theta_W}\ \frac{1}{\,
     {{\left( {Q^2} +
          {{{m_Z}}^2} \right)
         }^2}}
\nonumber \\
&\times& \sum_n\frac{\langle 0|{\cal O}_n|0\rangle}{m_c^3}
\int_{\frac{Q^2+4m_c^2}{s}}^1dx\,f_{g/N}\left(x,Q^2\right)
h_n\left(y,Q^2\right),
\label{eq:struct1}
\end{eqnarray}
where $s$ is the total invariant mass of the $\nu N$ system, $x$ is
the momentum fraction of the incoming gluon, $-Q^2$ is the momentum-squared
transferred from the leptonic system, $y=\frac{Q^2+4m^2}{sx}$, and
$f_{g/N}(x,Q^2)$ is a gluon distribution function in the nucleon.
The charmonium structure functions are given by
\begin{eqnarray}\label{eq:structf}
h_{\,^1\!S_0^{(8)}}\left(y,Q^2\right)&=&\left(g_V^c\right)^2 \,\times 6
\,{\frac{Q^2\, m_c^2}
    {{\left(Q^2 + 4m_c^2 \right) }^2} }\,(y^2-2y+2)
\nonumber\\[.1in]
h_{\,^3\!S_1^{(8)}}\left(y,Q^2\right)&=&\left(g_A^c\right)^2 \,\times 2
m_c^2\,\frac{Q^2(y^2-2y+2)+16(1-y)m_c^2}{\left(Q^2+4m_c^2\right)^2}
\nonumber\\h_{\,^3\!P_0^{(8)}}\left(y,Q^2\right)&=& {\left(g_V^c\right)^2
\,\times 2\,{Q^2}\,}
      {\,\frac{     {{\left(Q^2 + 12m_c^2 \right) }^2}}{
     {{\left( {Q^2} + 4m_c^2  \right) }^4}}}\,(y^2-2y+2)
\\[0.1in]
h_{\,^3\!P_1^{(8)}}\left(y,Q^2\right)&=&\left(g_V^c\right)^2 \,\times
4\,Q^4\,{ \frac{
   Q^2(y^2-2y+2) + 16(1-y)m_c^2
     }{{{\left( {Q^2} + 4m_c^2 \right) }^4}}}
\nonumber\\h_{\,^3\!P_2^{(8)}}\left(y,Q^2\right)&=&
     \left(g_V^c\right)^2\,\times \frac{4}{5}\,Q^2\,\biggl[
\frac{(y^2-2y+2) Q^4}{ {{\left(Q^2 + 4m_c^2 \right) }^2}}
\nonumber \\
&& ~~~~+~
{\frac{
      48(1-y)Q^2m_c^2+ 96(y^2-2y+2){{{m_c}}^4}
       }{ {{\left(Q^2 + 4m_c^2 \right) }^2}}} \biggr],
   \nonumber
\end{eqnarray}
where $g_A^c=\frac{1}{2}$ and
$g_V^c=\frac{1}{2}\left(1-\frac{8}{3}\sin^2\Theta_W\right)$
are the vector and axial couplings of the $c$-quark. Clearly, the coupling
constants favor the $\,^3S_1^{(8)}$ contribution, which is due to the 
large axial coupling (a similar contribution is, of course, absent in 
the case of $J/\psi$ lepto- and photoproduction).
Indeed a numerical estimate~\cite{Petrov:1999fm} shows that this matrix element
dominates the total cross section, and also the differential cross section 
unless $Q^2 \gg m_c^2$. At large $Q^2$, the relative $Q^4$ enhancement of 
the $P$-wave structure functions makes them dominant.
These structure functions should be incorporated in the specific Monte Carlo
generators built for the particular detector design.

\begin{table}
\caption{\label{table}Total cross sections for the $J/\psi$ production
in $\nu N\to J/\psi X$ for various incident neutrino energies.}
\begin{tabular}{ccccc}
$E_\nu[GeV]$   & 7.5  & 25    &  120  &   450  \\  \tableline
$\sigma[nb]$     & $7.8\times10^{-13}$&  $6.9\times10^{-10}$ &
$1.3\times10^{-8}$ &  $5.5\times10^{-8} $
\end{tabular}
\end{table}

An important question to address is the expected event rate of
$J/\psi$ production. Computing the total cross sections for 
the $J/\psi$ production (Table (\ref{table})), 
a simple calculation shows that currently
running neutrino experiments NOMAD and NuTeV could collect a 
few $J/\psi$ events (due to either low energy of the neutrino beam 
or particular detector configuration) and ``confirm'' the color octet 
mechanism. On the contrary, a neutrino experiment at the future
Muon Collider would collect about  $3\times 10^3$ events/year and
provide precise measurement of various NRQCD matrix elements and/or
the gluon distribution function.

\subsection{Neutrino factory $=$ charm factory?} 
It is clear from the preceding discussion that charm production plays
an important role in the studies of nucleon structure and electroweak
parameters. It is also important that with the estimated $10^8$ 
well-reconstructed charm events~\cite{Bigietal} \nuf~ is also an 
impressive {\it charm factory}. Charm physics is an important
complement to the $B$-physics program at $B$-factories 
(see, e.g.~\cite{BaBar}). Besides testing our understanding of
QCD effects in charmed particle decays, it also offers an
opportunity to look for the effects of new physics in rare decays
of charmed mesons, CP-violating asymmetries and $D \bar D$ mixing
studies, as the standard model background to this processes 
is tiny~\cite{AP}. 

It is interesting to see if \nuf~ has any advantages over 
the existing charm experiments.
One important advantage of $D \bar D$ mixing analysis 
performed at \nuf~ that is not available elsewhere involves 
initial $D$ flavor tagging. In particular,
$D^0$ mesons produced in {\it charged} current interactions receive 
an automatic initial flavor tag in the form of the final state lepton
charge. Correlation studies of the charges of the ``tag'' lepton and, 
say, lepton from the semileptonic charm decay would offer experimentally 
clean signatures of $D \bar D$ mixing.

\section{Rare processes} 
Neutrino-nucleon processes at low momentum transfer 
are sensitive to generic four-fermion contact terms produced by the
high energy neutral current interactions. 
These four-fermion interactions can be associated with 
supersymmetric theories with ${\cal R}$-parity nonconservation,
new vector bosons, quark compositness or even loop effects
associated with the new flavor-changing neutral current 
interactions~\cite{Bigietal}.
Consider, for instance, the low energy remnant of a generic high energy 
electron-quark neutral current interaction. It can be represented by
\begin{eqnarray} \label{nci}
{\cal L}_{NC} &=& 
\sum_q \Bigl[ \eta_{LL}^{eq}\left(\overline{e_L} \gamma_\mu e_L\right)
\left(\overline{q_L} \gamma^\mu q_L \right) + 
\eta_{RR}^{eq} \left(\overline{e_R}
\gamma_\mu e_R\right) \left( \overline{q_R}\gamma^\mu q_R\right) \nonumber\\
&& \quad {}+ \eta_{LR}^{eq} \left(\overline{e_L} \gamma_\mu e_L\right)
\left(\overline{q_R}\gamma^\mu q_R\right) + 
\eta_{RL}^{eq} \left(\overline{e_R} \gamma_\mu e_R\right)
\left(\overline{q_L} \gamma^\mu q_L \right) \Bigr] \,. 
\end{eqnarray}
A similar equation can be written for a direct neutrino-quark interactions.
One can use $SU(2)$ symmetry to relate $\nu$ and $e$ couplings
\begin{eqnarray}
\eta^{\nu u}_{LL} & = & \eta^{ed}_{LL}\; ,~
\eta^{\nu d}_{LL} = \eta^{eu}_{LL}\; ,
\nonumber \\
\eta^{\nu u}_{LR} & = & \eta^{eu}_{LR}\; ,~
\eta^{\nu d}_{LR} = \eta^{ed}_{LR}\; ,
\nonumber
\end{eqnarray}
so that $\nu N$ interactions can be used to constrain $\eta$'s of the 
Lagrangian of Eq.~(\ref{nci}). A particular example of a high-energy
model that leads to the low-energy Lagrangian of this type
is provided by ${\cal R}$-parity violating SUSY, where 
at low values of transferred momenta one can integrate out
heavy $\tilde e_{^L_R}^i$ and $\tilde d_{^L_R}^i$ to 
rewrite the Lagrangian in terms of local four-fermion 
interactions.
Assuming that the squarks of first two generations are degenerate and 
imposing $SU(2)$ symmetry constraints,
\begin{equation}
\eta_{LR}^{ed}= -{(\lambda'_{1j1})^2\over 2m_{\tilde u_L^j}^2} =
-{(\lambda'_{1j1})^2\over 2m_{\tilde d_L^j}^2} = \eta_{LR}^{\nu d}.
\end{equation}
Here $\lambda'_{ijk}$ is a parameter of the original $\rlap/{\cal R}$ 
SUSY Lagrangian.
Indeed, the best constraint on this coupling, 
$\eta_{LR}^{e d} < 0.07^{+0.24}_{-0.24}$ comes from the 
analysis of neutrino nucleon scattering 
experiments~\cite{Zeppenfeld:1998un}. Other new physics scenarios 
involve new heavy neutral leptons (models with $H^0_L-\nu_\mu$ 
mixing)~\cite{Gronau:1984ct} or new neutral gauge bosons like 
$Z'$ which appears in many superstring-motivated models.

\section{Conclusions}

New experiments utilizing high energy and intensity neutrino beams 
would offer a unique opportunity to perform new precision studies 
of QCD and electroweak interactions. New types of measurements, like
charmonium production and extractions of $V_{cb}$ and $V_{ub}$
CKM matrix elements, will become possible. Interesting new physics 
scenarios can also be explored. As a result, a high intensity neutrino 
facility could prove to be a very useful addition to the Muon Collider 
physics program.

\end{document}